\input{aipcheck}

\documentclass[
    ,final            
  ]
  {aipproc}

\layoutstyle{6x9}

\usepackage{amsmath}
\usepackage{amssymb}
\usepackage{wasysym}
\begin{document}

\title{Role of Bulk flow in Turbulent Convection}


\author{Mahendra K. Verma}{
  address={Department of Physics, Indian Institute of Technology, Kanpur, 208016, India}
}

\author{Ambrish Pandey}{
  address={Department of Physics, Indian Institute of Technology, Kanpur, 208016, India}
}

\author{Pankaj K. Mishra}{
  address={Laboratoire de Physique Statistique, Ecole Normale Superieure, Paris Cedex 05, France.}
}

\author{Mani Chandra}{
  address={Department of Astronomy, University of Illinois, Urbana-Champaign, USA}
}

\begin{abstract}
In this paper we present scaling of large-scale quantities like Pecl\'{e}t and Nusselt numbers, and the dissipation rates of kinetic energy and entropy.  Our arguments are based on the scaling of bulk quantities and earlier experimental and simulation results.  We also present the inertial-range properties of spectra and fluxes of kinetic energy and entropy.  
\end{abstract}

\maketitle
\section{1.~Introduction}

At first, we congratulate Prof. Abhijit Sen for his tremendous scientific achievements and for building Institute for Plasma Research.  He is one of the pioneers in nonlinear dynamics in India.  I thank him for his advise and encouragement to me throughout my career.  We wish him and his family a healthy and prosperous future.  Now we present some of our recent results on turbulent convection in his honour.

Thermal convection is an important phenomenon involving instability, pattern formation, chaos, and turbulence~\cite{Siggia:ARFM1994,Ahlers:RMP2009}.  Physics of convective turbulence differs significantly from that of homogenous isotropic turbulence due to the presence of the thermal plates and the buoyancy. Thermal plumes are generated in the boundary layer near the plate, after which they are transported to the bulk of the fluid.  

Scaling of large-scale quantities like Pecl\'{e}t and  Nusselt numbers, energy spectra, etc. are of interest in convective turbulence.  They have been modelled using the properties of bulk and boundary layers~\cite{Kraichnan:PF1962,Shraiman:PRA1990,Cioni:JFM1997,Grossmann:JFM2000}, with the work of Grossmann and Lohse~\cite{Grossmann:JFM2000} (GL) being the most recent one.  The predictions of GL theory are in close agreement with the results of experiments and numerical simulations of Rayleigh B\'{e}nard convection (RBC).   In this paper we derive the properties of large-scale quantities, as well as the spectra of energy and entropy, by applying scaling arguments to the bulk flow.  We assume that the turbulence in the bulk is fed by the thermal plumes generated in the boundary layer.    We will show below that significant conclusions can be drawn by focussing on the bulk turbulence itself.   Our results based on scaling arguments are consistent with the GL theory, and are in reasonable agreement with earlier experimental and numerical results.     

The outline of the paper is as follows: In Sec.~2 we discuss the governing equations and numerical details. The scalings of large scale quantities, e.g., large-scale velocity and large-scale temperature fluctuations for different ranges of the control parameters are presented in Sec.~3.  The dependence of Nusselt number and dissipation rates on $\mathrm{Ra}$ and $\mathrm{Pr}$ are derived in Sec.~4 and 5 respectively.   In Sec.~6, we present the inertial-range energy and entropy spectra. In Sec.~7, we briefly describe the role of bulk motion on the reversals of large scale velocity in RBC.  We present summary and conclusions in Sec.~8.

\section{2.~Governing Equations}
Rayleigh-B\'{e}nard convection (RBC) is an idealised version of convective flow in which a fluid is placed between two plates.  The equations for RBC under Boussinesq approximation are 
\begin{eqnarray}
\partial_{t}\mathbf{u} + (\mathbf{u} \cdot \nabla)\mathbf{u} & = & -\frac{\nabla \sigma}{{\rho}_0} + \alpha g \theta \hat{z} + \nu{\nabla}^2 \mathbf{u}, \label{eq:u} \\
\partial_{t}\theta + (\mathbf{u} \cdot \nabla)\theta & = & \frac{\Delta}{d} u_{z} + \kappa{\nabla}^{2}\theta, \label{eq:theta} \\
\nabla \cdot \mathbf{u} & = & 0 \label{eq:cont},
\end{eqnarray}
where, $\mathbf{u} = (u_x,u_y,u_z)$ is the velocity field, $\sigma$ and $\theta$ are, respectively, the  pressure and temperature fluctuations from the steady conduction state (Total temperature $T = T_c + \theta$, with $T_c$ as the conduction temperature profile), $d$ is the distance between the plates, and $\Delta$ is the temperature difference. Also, $\hat{z}$ is the buoyancy direction, ${\rho}_0$ is the mean density of fluid, $g$ is the acceleration due to gravity, and $\alpha$, $\nu$, and $\kappa$ are the thermal expansion coefficient, the  kinematic viscosity, and the thermal diffusivity of fluid, respectively.   

The important nondimensional parameters of RBC  are the Reynolds number ${\mathrm{Re}}$,  the Pecl\'{e}t number ${\mathrm{Pe}}$, the Rayleigh number ${\mathrm{Ra}}$, the Prandtl number ${\mathrm{Pr}}$, and the Nusselt number ${\mathrm{Nu}}$.  They are derived based on dimensional analysis as 
\begin{eqnarray}
\mathrm{Re} & = &\frac{{\bf u}\cdot \nabla {\bf u}}{\nu \nabla^2 {\bf u}} = \frac{U d}{\nu}, \\
\mathrm{Pe} & = & \frac{{\bf u}\cdot \nabla \theta}{\kappa \nabla^2 \theta} = \frac{U d}{\kappa}, \\
\mathrm{Ra} &= &\frac{\alpha g \theta}{\nu \nabla^2 {\bf u}} = \frac{\alpha g \Delta d^3}{\nu \kappa}, \\
\mathrm{Pr} & = & \frac{\nu}{\kappa}, \\
\mathrm{Nu} &= &\frac{\mathrm{Total~heat~flux}} {\mathrm{Conductive~heat~flux}}= \frac{\langle -\kappa \nabla T +u_z T \rangle^{xy}}{\langle -\kappa \nabla T \rangle^{xy}}. 
\label{eq:Nu_definition}
\end{eqnarray}
For the Nusselt number computation, the average of the heat flux is performed over the horizontal planes. Under steady state, the flux over each of the horizontal planes must be constant.    Hence, we can also compute the Nusselt number by volume averaging, i.e.,
\begin{equation}
\mathrm{Nu} = \frac{\langle -\kappa \nabla T +u_z T \rangle^{xyz}}{\langle -\kappa \nabla T \rangle^{xyz}} = 1 + \left \langle \frac{u_z d}{\kappa} \frac{\theta}{\Delta} \right \rangle^{xyz} = 1 +\langle u_z' \theta' \rangle^{xyz},   \label{eq:Nu_definition2}
\end{equation}
where $u_z'=u_z d/\kappa$ and $\theta'=\theta/\Delta$.   It is also important to note that $\mathrm{Pe = RePr}$.

The above set of equations are nonlinear, and they do not have simple solutions.  However, theoretical models~\cite{Kraichnan:PF1962,Shraiman:PRA1990,Cioni:JFM1997,Grossmann:JFM2000} based on scaling arguments are able to explain the experimental and numerical results reasonably successfully.   In this paper we present another set of scaling arguments that relate the response parameters $\mathrm{Pe}, \mathrm{Re}$, and $\mathrm{Nu}$ with the control parameters $\mathrm{Ra}$ and $\mathrm{Pr}$.  We also compare our results with earlier experimental and simulation results.   We point out that the scaling exponents of bulk quantities for both free-slip and no-slip boundary conditions are quite close.  However the prefactors for $\mathrm{Pe}, \mathrm{Re}$, and $\mathrm{Nu}$  for the free-slip walls are larger compared to those for no-slip walls, which is due to a smaller frictional force for the free-slip walls compared to the no-slip walls.

We compare our model predictions with earlier experimental and numerical results.  For
some parameters, specially for $\mathrm{Pr}=0$ and $\infty$, we too have performed numerical simulations for free-slip and no-slip walls (top and bottom ones).  We assume periodic boundary conditions along the horizontal directions.  We use ``Tarang'' developed by Verma \textit{et al.}~\cite{Verma:sub2012} for free-slip boundary conditions and ``Nek5000'' for no-slip boundary conditions.   For further details of the simulations, see Mishra and Verma~\cite{Mishra:PRE2010}.

In the following sections we will describe  the scaling of large-scale quantities, as well as that of energy and entropy spectra, as a function of $\mathrm{Ra}$ and $\mathrm{Pr}$.

\section{3.~Scaling of large-scale quantities}
In this section we describe the scaling of the large-scale velocity $U_L$ and temperature fluctuation $\theta_L$.   One of the generic features observed in all our numerical simulations are the finite amplitude of Fourier $\hat{\theta}(0,0,2n)$ modes~\cite{Mishra:PRE2010}, where the three indices indicate wavenumber components ($k_x,k_y,k_z$).  We observe that 
\begin{equation}
\hat{\theta}(0,0,2n) \approx -\frac{\Delta}{2 n \pi}.
\end{equation}
The above scaling occurs due to the fact that the entropy transfer from the mode $\hat{\theta}(n,0,n)$ to the mode $\hat{\theta}(0,0,2n)$ is approximately equal to entropy production due to the $\hat{u_z}(n,0,n)$ mode, specially for lower $n$'s ($n=1,2,3$).  A detailed derivation is given in Mishra and Verma~\cite{Mishra:PRE2010}.

\begin{figure}
\includegraphics[scale=0.5]{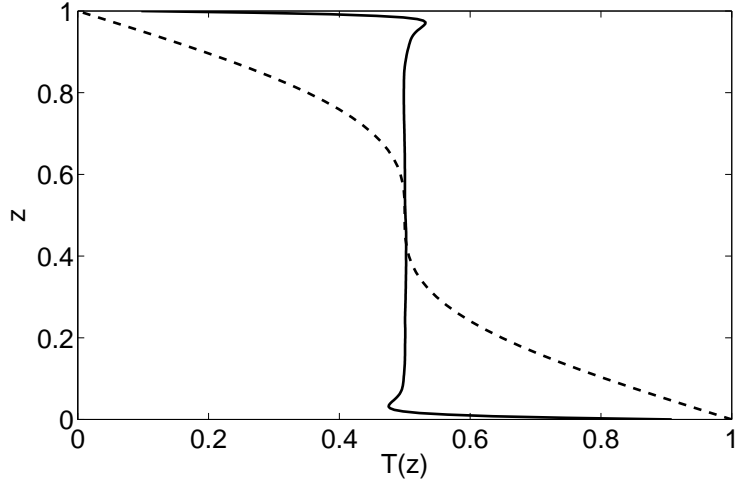}
\caption{Variation of the averaged temperature profile (solid line)  along the vertical direction for Pr = $\infty$, Ra = $10^8$ (simulation grid size=$512^3$).  The averaged temperature remains constant in the bulk, while it displays sharp gradient near the top and bottom plates. Dashed line represents $T_c+\hat{\theta}(0,0,2)$, where $T_c$ is the conduction profile.}
\label{fig:temp}
\end{figure}

The mode $\hat{\theta}(0,0,2n)$ has an important consequence on the vertical profile of temperature.  The averaged temperature over horizontal planes drops sharply in the boundary layer, and it is approximately constant in the bulk.  As shown in Fig.~\ref{fig:temp}, $\hat{\theta}(0,0,2)$ contributes significantly to the temperature drop near the plates.  These results demonstrate the important role played by the $\hat{\theta}(0,0,2n)$ modes in turbulent convection.  

It is also important to note that $\hat{\theta}(0,0,2n)$ do not contribute to buoyancy.  The corresponding $\hat{u_z}(0,0,2n) = 0$ because the net mass flux across any horizontal plane must be zero.  As a result,
\begin{equation}
{\bf u} \cdot \nabla {\bf u} \approx \alpha g \theta_{\mathrm{res}},
\end{equation}
where $\theta_{\mathrm{res}}^2 = \theta^2 - \sum_n |\hat{\theta}(0,0,2n)|^2$ (here "res" stands for residual).  Using these relations we can write equations to estimate $U_L$ and $\theta_L$ that are
\begin{eqnarray}
c_1 \frac{U_L^2}{d} & = & \alpha g \theta_{\mathrm{res}} + c_2 \nu \frac{U_L}{d^2} \label{eq:UL}  \\
c_3 \frac{U_L \theta_L}{d}  & = &   \frac{\Delta}{d} U_L + c_4 \kappa \frac{\theta_L}{d^2}
\end{eqnarray}
where $c_1,c_2,c_3,c_4$ are constants to be determined using the data obtained in experiments and numerical simulations.  These constants may be weak functions of system parameters like Prandtl number, aspect ratio, etc.  In the following discussion we derive
scaling of $U_L$ and $\theta_L$  in the limiting cases: (1) $\mathrm{Re} \gg 1; \mathrm{Pe} \gg 1$, (2) $\mathrm{Re} \gg 1; \mathrm{Pe} \ll 1$, and (3) $\mathrm{Re} \ll 1; \mathrm{Pe} \gg 1$.  The thickness of the boundary layers is reasonably small for Case 1.   However, the boundary layers cover most of the bulk for the cases 2 and 3, thus making the scaling of the bulk flow very relevant, specially for the limiting cases.

\subsection{Case 1: $\mathrm{Re} \gg 1; \mathrm{Pe} \gg 1$}
In this regime, the nonlinear terms of both Eqs.~(\ref{eq:u},\ref{eq:theta}) are much larger than the diffusive terms.  Therefore, by matching the most dominant terms of the equations, we obtain
\begin{eqnarray}
{\bf u} \cdot \nabla \theta \approx \frac{\Delta}{d} u_z, \\
{\bf u} \cdot \nabla {\bf u} \approx \alpha g \theta_{\mathrm{res}}.
\end{eqnarray}
Numerical simulations~\cite{Mishra:PRE2010} reveal that $\theta_{\mathrm{res}} \sim \hat{\theta}(0,0,2) \sim \Delta$.  Therefore,
\begin{eqnarray}
\theta_L \approx \Delta, \\
\mathrm{Pe} = \frac{U_L d}{\kappa} \approx \sqrt{\mathrm{Ra Pr}}. \label{eq:PeRe_large}
\end{eqnarray}
The proportionality constant depends on the boundary condition.  Experiments and numerical simulations limit the constant to lie between 0.1 to 0.4.  For this case, Eq.~(\ref{eq:PeRe_large}) implies that $\mathrm{Ra Pr} \gg 1$.

\subsection{Case 2: $\mathrm{Re} \gg 1; \mathrm{Pe} \ll 1$}
Since $\mathrm{Pe} \ll 1$, the term $(\Delta/d) u_z$ matches with the diffusive term in Eq.~(\ref{eq:theta}).   The condition $\mathrm{Re} \gg 1$ implies that in Eq.~(\ref{eq:u}), the nonlinear term matches with the buoyancy term.  Therefore,
\begin{eqnarray}
\frac{\Delta}{d} u_z \approx \kappa \nabla^2 \theta_{\mathrm{res}},
\label{eq:Przer0_uz_theta}\\
{\bf u} \cdot \nabla {\bf u} \approx \alpha g \theta_{\mathrm{res}},
\end{eqnarray} 
or,
\begin{eqnarray}
\theta_{\mathrm{res}} \approx \mathrm{Ra Pr}, \\
\mathrm{Pe} = \frac{U_L d}{\kappa} \approx \mathrm{Ra Pr}.
\end{eqnarray} 
Since $\mathrm{Pe} = \mathrm{Re Pr} $,  we obtain
\begin{equation}
\mathrm{Re} = \mathrm{Ra}.
\end{equation}
Note that $\theta_L$ is dominated by $\hat{\theta}(0,0,2n)$ modes, hence 
\begin{equation}
\theta_L \approx \Delta.
\end{equation}
The condition $\mathrm{Pe} \ll 1$ implies that $\mathrm{Ra Pr} \ll 1$ for this case.

\subsection{Case 3: $\mathrm{Re} \ll 1; \mathrm{Pe} \gg1$}
In this case, we ignore the nonlinear term of Eq.~(\ref{eq:u}), but not in the temperature equation, which yields
\begin{eqnarray}
{\bf u} \cdot \nabla \theta  & \approx & \frac{\Delta}{d} u_z,  \\
\nu \nabla^2 \mathbf{u} & \approx & \alpha g \left[\sum_\mathbf{k} |\hat{\theta}(\mathbf k)|^2 \frac{k^2_\perp}{k^6}\right]^{1/2},
\end{eqnarray}
where $\left[\sum_\mathbf k |\hat{\theta}(\mathbf k)|^2 \frac{k^2_\perp}{k^6}\right]^{1/2}$ appears due to the significant contributions arising from the pressure gradient term~\cite{Pandey:Preprint}. Therefore,
\begin{equation}
\theta_L \approx \Delta.
\end{equation}
Numerical simulations by Pandey \textit{et al.}~\cite{Pandey:Preprint} indicate that $\theta_{\mathrm{res}} \ll |\hat{\theta}(0,0,2)|$ and 
\begin{equation}
 \theta_{\mathrm{res}} \approx \mathrm{Ra}^{-\delta}  \Delta,
 \end{equation}
with $\delta \approx 0.15$. Pandey \textit{et al.}~\cite{Pandey:Preprint} also show that $\left[\sum_\mathbf k |\hat{\theta}(\mathbf k)|^2 \frac{k^2_\perp}{k^6}\right]^{1/2} \sim \Delta  \mathrm{Ra}^{-\zeta}$ with $\zeta \approx 0.38$. As a result,
\begin{equation}
\mathrm{Pe} \approx \mathrm{Ra}^{1-\zeta}  \approx \mathrm{Ra}^{0.62}. 
\end{equation}

\begin{figure}
\includegraphics[scale=0.45]{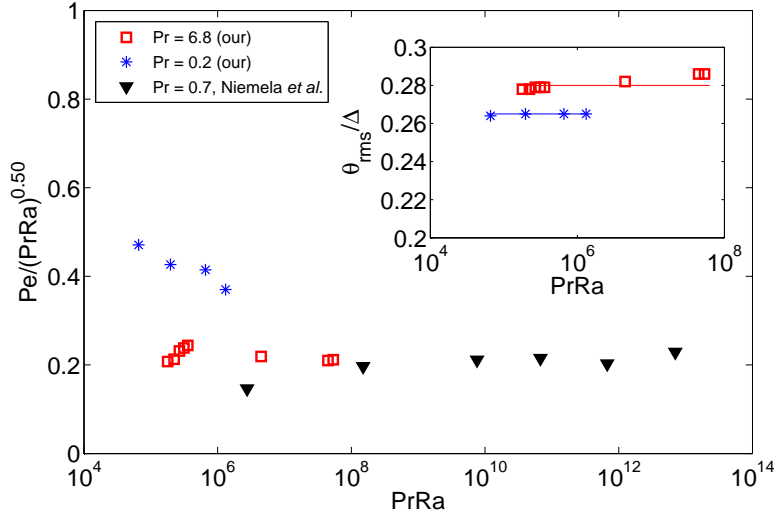}
\caption{Normalized Peclet number (Pe/(PrRa)$^{0.5}$) as a function of PrRa: simulation data for Pr=6.8 (red box) and Pr=0.2 (blue star), and experimental data for Pr=0.7 ~\cite{Niemela:JFM2001}. The inset shows that  the normalized large-scale temperature is constant with relative to PrRa. [Adopted from Verma \textit{et al.}~\cite{Verma:PRE2012}].}
\label{fig:pe_pr}
\end{figure}
 
\begin{figure}
\includegraphics[scale=0.45]{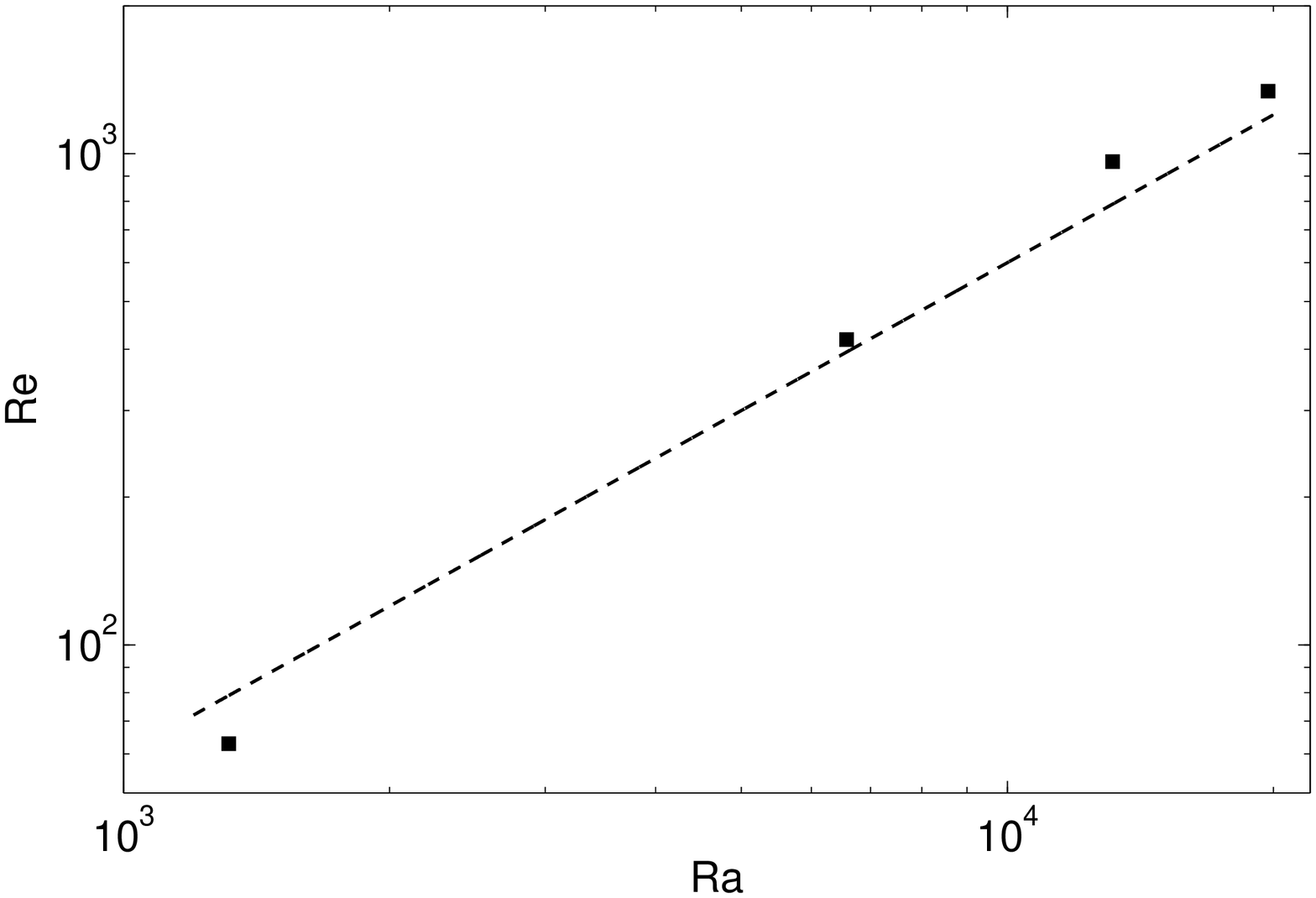}
\caption{The Reynolds number Re as a function of the Rayleigh number Ra for Pr = 0. The best fit (dotted curve) to the data yields the scaling relationship as Re = $(0.06 \pm 0.23)\mathrm{Ra}^{1.01 \pm 0.39}$.}
\label{fig:Re_Ra}
\end{figure}

Most of the experiments and numerical simulations of convective turbulence correspond to case 1 for which $\mathrm{Re} \gg 1$ and  $\mathrm{Pe} \gg 1$.  For this case, the scaling arguments are in excellent agreement with the experimental~\cite{Niemela:JFM2001} and numerical simulations~\cite{Verma:PRE2012} as exhibited by Fig.~\ref{fig:pe_pr}. The above arguments are also in general agreement with scaling analysis of Grossmann and Lohse~\cite{Grossmann:JFM2000}.  So far, no numerical or laboratory experiments have been performed corresponding to case 2, except for $\mathrm{Pr}=0$ that exhibits $\mathrm{Re} = \mathrm{Ra}$ (see Fig.~\ref{fig:Re_Ra}).  Regarding case 3, our scaling arguments are in good agreement with recent simulations by  Verzicco and Camussi~\cite{Verzicco:JFM1999}, Verzicco and Sreenivasan~\cite{Verzicco:JFM2008}, and Pandey \textit{et al.}~\cite{Pandey:Preprint}.

In the next section, we will discuss Nusselt number scaling.
 
 
 \section{4.~Scaling of Nusselt number}
 
 Using Eq.~(\ref{eq:Nu_definition2}), 
\begin{equation}
\mathrm{Nu}-1 = \langle u_z' \theta'_{\mathrm{res}} \rangle = C^{u\theta}(\mathrm{Ra Pr})  \langle  u_z'^2  \rangle^{1/2}_V \langle\theta'^2_{\mathrm{res}}  \rangle_V^{1/2},
\end{equation}
where
\begin{equation}
C_{u\theta}(\mathrm{Ra Pr}) = \left \langle \frac{\langle u_z' \theta'_{\mathrm{res}}  \rangle_V} {  \langle u_z'^2  \rangle_V^{1/2}   \langle  \theta'^2_{\mathrm{res}}   \rangle_V^{1/2}} \right \rangle_t.
\end{equation}
Here $V$ and $t$ stand for the volume and temporal averages respectively.
For case 1, which is the most relevant one for experiments, we observe through numerical simulations that for $\mathrm{Pr} \approx 1$,
\begin{equation}
C_{u\theta}(\mathrm{Ra Pr}) \sim (\mathrm{Ra Pr})^{-0.2}.
\end{equation}
at least up to $\mathrm{Ra} \approx10^8$.  Using the results of  the earlier section, $(u_z')_L \approx (\mathrm{Ra Pr})^{1/2}$.  Therefore,
\begin{equation}
\mathrm{Nu}  \approx  (\mathrm{Ra Pr})^{1/2-0.2}  \approx ( \mathrm{Ra Pr})^{0.3},
\end{equation}
which is observed in  experiments ~\cite{Cioni:JFM1997, Castaing:JFM1989, Niemela:Nature2000, Glazier:Nature1999, Urban:PRL2011, Funfschilling:PRL2009} and numerical simulations ~\cite{Verzicco:JFM1999, Verzicco:JFM2008, Kerr:JFM2000, Silano:JFM2010, Stevens:JFM2010, Stevens:JFM2011}  up to $\mathrm{Ra} = 10^{14}$ or so.  Note that the parameters for the above simulations and experiments correspond to Case 1 for which $\mathrm{Re} \gg 1$ and $\mathrm{Pe} \gg 1$.

In a recent experiment, He \textit{et al.}~\cite{He:PRL2012} report an increase of the Nusselt number exponent to approximately 0.38 near $\mathrm{Ra}_{\mathrm{tr}} = 5 \times 10^{14}$.  He \textit{et al.}~\cite{He:PRL2012} attribute the above increase to the onset of the ``ultimate regime'' predicted by Kraichnan [1962].  Interestingly, the transitional Rayleigh number $\mathrm{Ra}_{\mathrm{tr}}$ corresponds to $\mathrm{Re} \approx \mathrm{Pe} \approx 0.1\times \sqrt{\mathrm{Ra}_{\mathrm{tr}}} \approx 4 \times 10^6$, which is close to the transitional Reynolds number for the emergence of the turbulent boundary layer in the flow over a flat plate, as well as for the flow past a cylinder.  Ahlers \textit{et al.}~\cite{Ahlers:PRL2012} reported a logarithmic profile for the temperature above $\mathrm{Ra} = \mathrm{Ra}_{\mathrm{tr}}$, which is in general agreement with the logarithmic profile for the velocity in the turbulent boundary layer for the flow past a flat plate.  This result appears to indicate a birth of a turbulent boundary layer, as well as destruction of the ``large-scale circulation'' beyond $\mathrm{Ra} = \mathrm{Ra}_{\mathrm{tr}}$.   The disappearance of the large-scale circulation beyond $\mathrm{Ra} = \mathrm{Ra}_{\mathrm{tr}}$ has been reported by Ahlers \textit{et al.}~\cite{Ahlers:PRL2012}.  Chavanne {\em et al.}~\cite{Chavanne:PRL1997} and Roche {\em et al.}~\cite{Roche:NJP2010} report the onset of ``ultimate regime'' near $\mathrm{Ra} \approx 10^{12}$ for non-smooth plates.  Their results are consistent with the fact that roughness of the plate can reduce the onset of turbulence in the boundary layer.

\begin{figure}
\includegraphics[scale=0.45]{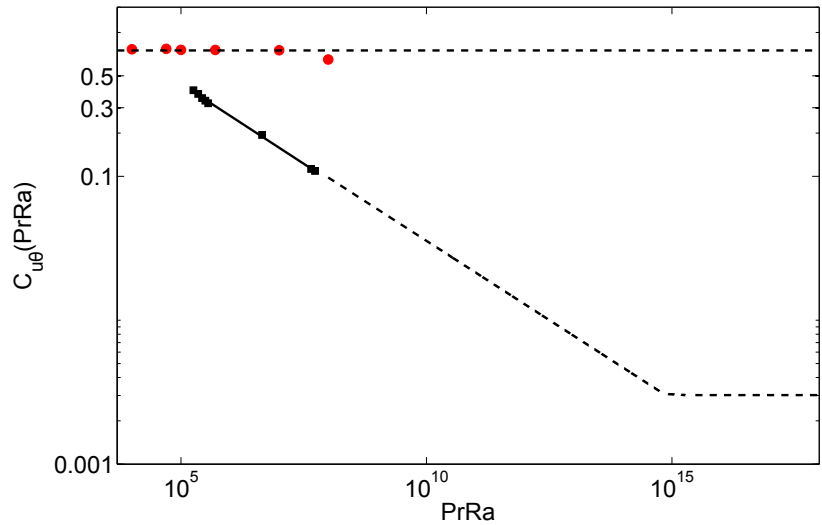}
\caption{The red circles with extended chained line depict constancy of the normalized correlation function $C_{u\theta}$ for the periodic box (Pr = 1). Assuming that the convective turbulence becomes fully-developed for very large Ra, we conjecture that the normalized correlation function would become a constant in the ultimate regime. The second curve is the correlation function for Pr = 6.8 along with extended Ra$^{-0.22}$ for moderately large Ra, and then a constant for the ultimate regime after some transitional Rayleigh number~\cite{Ahlers:PRL2012}.}
\label{fig:cos_zeta}
\end{figure}

The variation of $C_{u\theta} \sim (\mathrm{Ra Pr})^{-0.2}$ appears to be a key ingredient for the deviation from Kraichnan's prediction that $\mathrm{Nu} \sim \mathrm{Ra}^{1/2}$ for the ``ultimate regime''.  It is important to note that the correlation function $C_{u\theta}(\mathrm{Ra Pr})$  should flatten in the ultimate regime as indicated in Fig.~\ref{fig:cos_zeta}.
The change-over is expected to take place near $\mathrm{Ra} = \mathrm{Ra}_{\mathrm{tr}} \approx 5 \times 10^{14}$.

For the second case with $\mathrm{Re} \gg 1$ and $\mathrm{Pe} \ll 1$, the large-scale Fourier modes $\hat{u_z}({\bf k})$ and $\hat{\theta}_{\mathrm{res}} ({\bf k})$ are in phase since $\hat{u_z}({\bf k}) = -(\kappa k^2 d/\Delta)\hat{\theta}_{\mathrm{res}} ({\bf k})$ [see Eq.~(\ref{eq:Przer0_uz_theta})].   Therefore, $C_{u\theta}(\mathrm{Ra Pr}) \approx 1$.  Also, $(u_z')_L \approx (\theta'_{\mathrm{res}} )_L \approx \mathrm{Ra Pr}$.  Hence 
\begin{equation}
\mathrm{Nu}-1  \approx  (\mathrm{Ra Pr})^2.
\end{equation}
Since $\mathrm{Ra Pr} \rightarrow 0$, $\mathrm{Nu} \rightarrow 1$ for this case, indicating that the convective heat transport is negligible, consistent with the results of Kraichnan~\cite{Kraichnan:PF1962}.

For case 3 too, $u_z'({\bf k}) \propto \theta'({\bf k})$, but overall $u_z'$ is not proportional to $\theta'$ due to the summation.  Consequently $C_{u\theta}(\mathrm{Ra Pr}) \approx Ra^{-0.15}$. Using the scaling for $U_L$ and $\theta_{\mathrm{res}}$,  we obtain
\begin{equation}
\mathrm{Nu}-1  \approx  \mathrm{Nu} \approx C_{u\theta}(\mathrm{Ra Pr}) \langle  u_z'^2  \rangle^{1/2}_V \langle\theta'^2_{\mathrm{res}} \rangle_V^{1/2} \approx \mathrm{Ra}^{\zeta - \delta - 0.15} \approx \mathrm{Ra}^{0.32} .
\end{equation}
Note that $\delta \approx 0.15$ [see Eq.~(27)].

\begin{figure}
\includegraphics[scale=0.45]{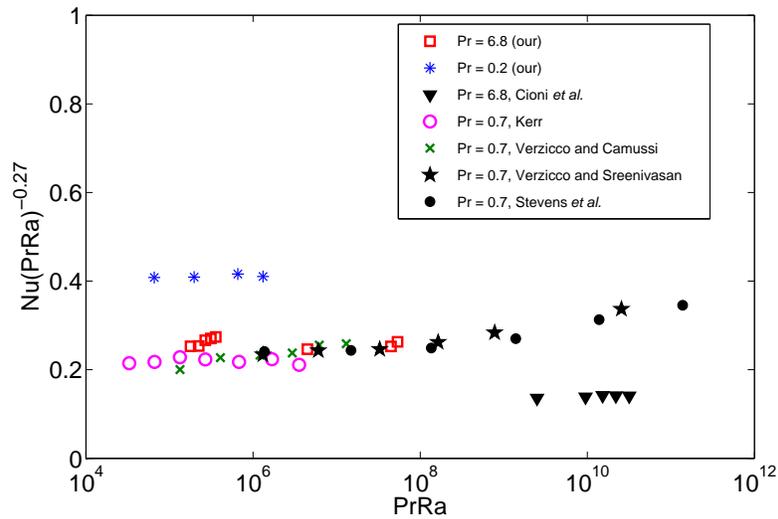}
\caption{Normalized Nusselt number (Nu/(PrRa)$^{0.27}$) as a function of PrRa. Red boxes and blue stars are the numerical simulation data respectively for Pr = 6.8 and Pr = 0.2 from Verma \textit{et al.}~\cite{Verma:PRE2012}, and inverted black triangles are the experimental data for Pr = 6.8 from Cioni \textit{et al.}~\cite{Cioni:JFM1997}. Pink circles (Kerr~\cite{Kerr:JFM1996}), green crosses (Verzicco and Camussi~\cite{Verzicco:JFM1999}), black stars (Verzicco and Sreenivasan~\cite{Verzicco:JFM2008}), and black dots (Stevens \textit{et al.}~\cite{Stevens:JFM2010}) are the numerical data for Pr = 0.7. [Adopted from Verma \textit{et al.}~\cite{Verma:PRE2012}].}
\label{fig:nu_pr}
\end{figure}

No experimental or numerical results have been reported for case 2. For case 3, however, our scaling arguments are in good agreement with numerical results of Pandey \textit{et al.}~\cite{Pandey:Preprint}.

In the next section we discuss the scaling of viscous dissipation rate $\epsilon_u$ and the entropy dissipation rate $\epsilon_\theta$.

 \section{5.~Scaling of dissipation rates}
 The two exact relations connecting the Nusselt number with the viscous dissipation rate $\epsilon_u$  and entropy dissipation rate $\epsilon_\theta$ are
 \begin{eqnarray}
\mathrm{Nu-1} & = & \frac{\mathrm{Pr}^2 d^4 \epsilon_u}{\nu^3 \mathrm{Ra}}, \\
 \mathrm{Nu}  & = & \frac{\epsilon_\theta d^2}{\kappa \Delta^2}. 	
\end{eqnarray}
The above equations can be rewritten as
\begin{eqnarray}
\mathrm{Nu-1} & = & \frac{\mathrm{Pe}^3}{ \mathrm{Ra Pr}} C_{\epsilon_u},
\label{eq:Nu_minus1} \\
 \mathrm{Nu}  & = &  \mathrm{Pe} C_{\epsilon_\theta}, \label{eq:Nu}
\end{eqnarray}
 where $C_{\epsilon_u} = \epsilon_u/(U_L^3/d)$ and $C_{\epsilon_\theta} = \epsilon_\theta/(U_L \theta_L^2/d)$. 
 
\begin{figure}
\includegraphics[scale=0.3]{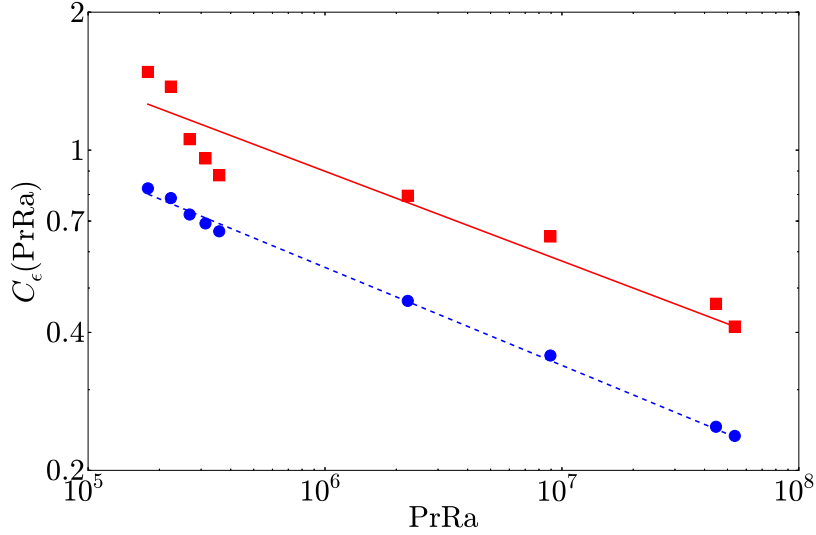}
\caption{Dissipation-rate correlation ($C_{\epsilon}$) as a function of PrRa for Pr = 6.8. Both viscous-dissipation correlation (red boxes) and thermal-dissipation correlation (blue circles) decreases with increase of Ra. The best fit to the data shows the scalings $C_{\epsilon_u} \sim (\mathrm{PrRa})^{-0.20}$ (red line) for viscous dissipation correlation and $C_{\epsilon_{\theta}} \sim (\mathrm{PrRa})^{-0.21}$ (blue dotted line) for thermal dissipation correlation.} 

\label{fig:corr_diss}
\end{figure}

 For case 1 ($\mathrm{Re} \gg 1, \mathrm{Pe} \gg 1$), our numerical data indicates that $C_{\epsilon_u} \approx C_{\epsilon_\theta} \approx (\mathrm{Ra Pr})^{-0.2}$ yielding $\mathrm{Nu} \approx (\mathrm{Ra Pr})^{0.3}$ (see Fig.~\ref{fig:corr_diss}).   For case 2, the flow is Kolmogorov-like, hence $C_{\epsilon_u} \approx 1$ and $\mathrm{Pe = Ra Pr}$.  Therefore, Eq.~(\ref{eq:Nu_minus1}) yields $\mathrm{Nu-1} \approx (\mathrm{Ra Pr})^2$.   Using $\epsilon_\theta = \kappa \theta_L^2/d^2$ or $C_{\epsilon_\theta} = 1/\mathrm{Pe}$, we can deduce using Eq.~(\ref{eq:Nu})  that $\mathrm{Nu} \approx 1$.  These two deductions are consistent with each other.
 
 For case 3, Pandey \textit{et al.}~\cite{Pandey:Preprint} show that $C_{\epsilon_\theta} \approx  \mathrm{Ra}^{-a}$ with $a \approx 0.29$, hence 
\begin{equation}
 \mathrm{Nu = Pe} C_{\epsilon_\theta}  \approx \mathrm{Ra}^{1-\zeta-a} \approx \mathrm{Ra}^{0.33},
 \end{equation}
since $\zeta \approx 0.38$.  
  
The above results are consistent with the $\mathrm{Nu}$ scaling predicted using the large-scale quantities. 

\section{6.~Energy spectra and fluxes}
Buoyancy acts at all length scales, hence, Kolmogorov's theory in which turbulence is forced at large scales may not hold for convective turbulence.  We need to examine the phenomenology of convective turbulence carefully.  We perform our analysis for the three case discussed earlier.
\subsection{Case 1:$\mathrm{Pe} \gg 1; \mathrm{Re} \gg 1$}

\begin{figure}
\includegraphics[scale=0.4]{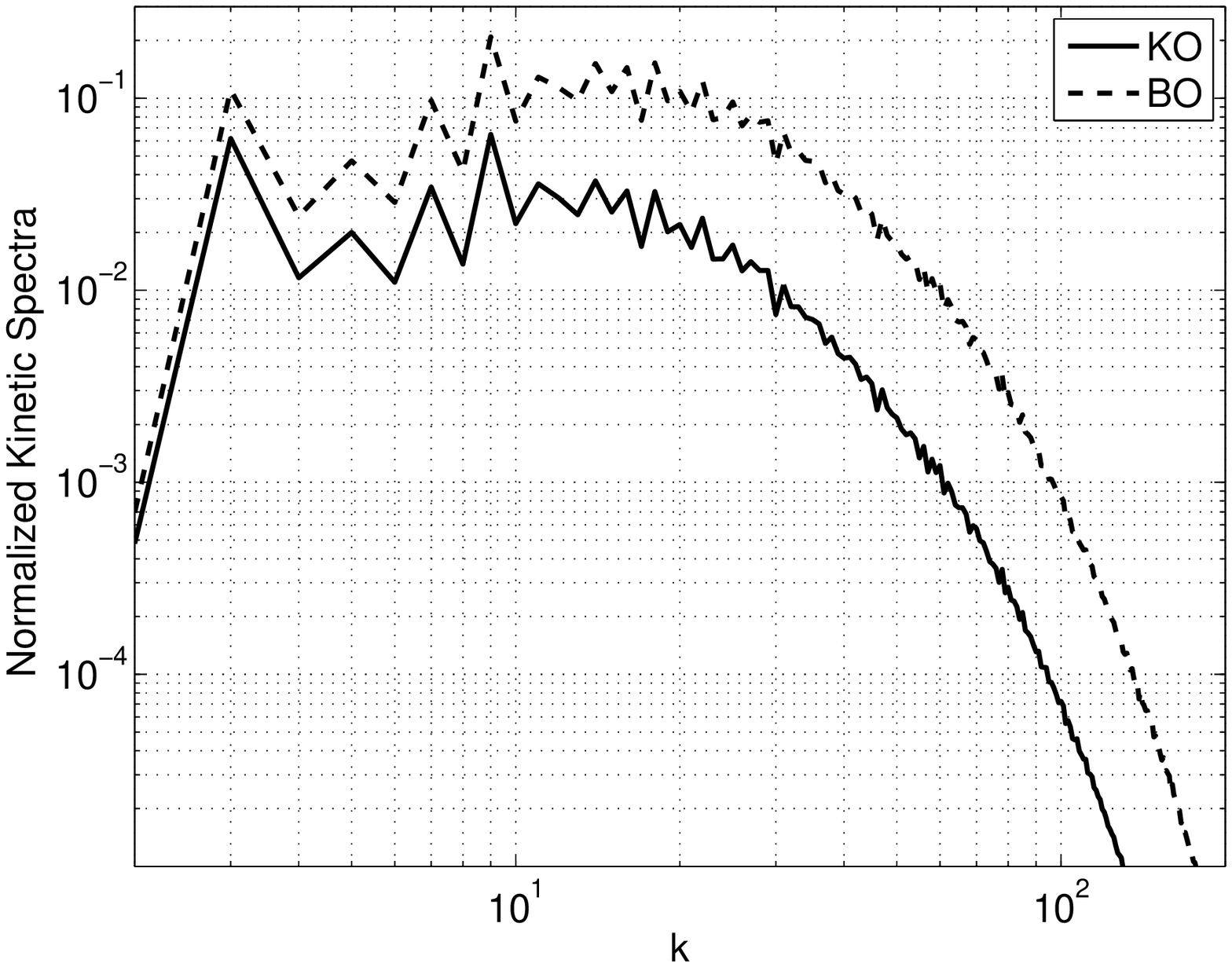}
\includegraphics[scale=0.4]{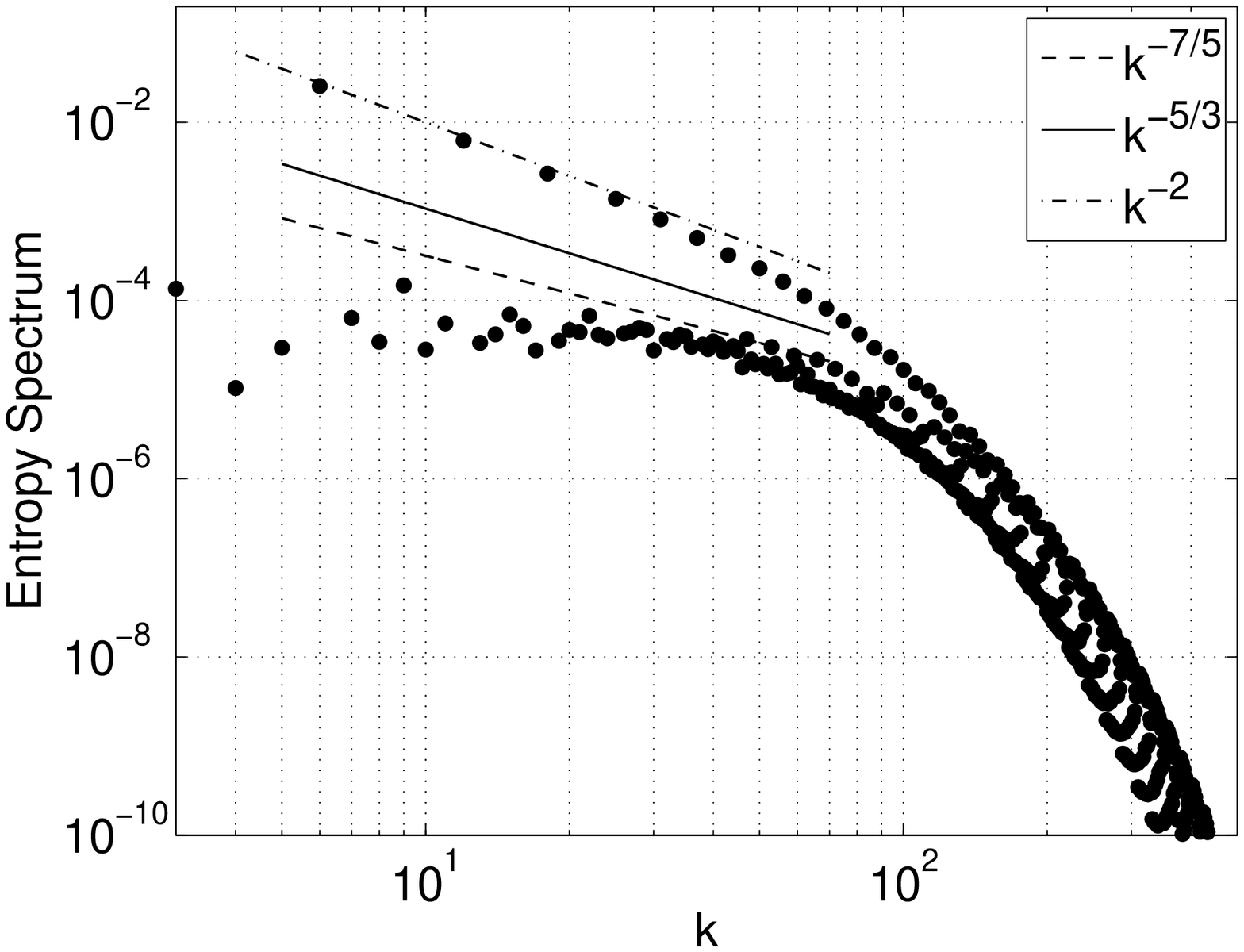}
\caption{Plot of the normalized kinetic spectra (left) and the entropy spectrum (right) for Pr = 6.8 and Ra = $6.6 \times 10^6$.  The numerical data is not consistent with either of the two models. However, the upper branch of the entropy spectrum fits well with $k^{-2}$ scaling. [Adopted from Mishra and Verma~\cite{Mishra:PRE2010}].}
\label{fig:spectrum_pr6p8}
\end{figure}

When buoyancy is important, and the buoyancy term is matched with the nonlinear term, then
\begin{equation}
u_l^3/l \approx \alpha g \theta_l u_l C_{u\theta}.
\end{equation}
Assuming constant entropy cascade for $\theta$, we obtain
\begin{equation}
\epsilon_\theta \approx (u_L \theta^2_L/d) C_{u\theta} \approx (u_l \theta^2_l/l) C_{u\theta}.
\end{equation}
Using $C_{u\theta} \approx (\mathrm{Ra Pr})^{-0.2}$, we deduce
\begin{eqnarray}
\theta_k & \approx & \Delta (\mathrm{RaPr})^{0.04} (kd)^{-1/5}, \\
u_k & \approx &\frac{\kappa}{d} (\mathrm{RaPr})^{0.42} (kd)^{-3/5}, 
\end{eqnarray}
or,
\begin{eqnarray}
E_u(k) & \approx & d \left( \frac{\kappa}{d} \right)^2 (\mathrm{RaPr})^{0.84} (kd)^{-11/5}, \\
E_\theta(k) & \approx &\Delta^2 (\mathrm{RaPr})^{-0.4} (kd)^{-7/5}, \\
\Pi_u(k) &  \approx & \frac{1}{d} \left(\frac{\kappa}{d} \right)^3 (\mathrm{RaPr})^{1.26} (k d)^{-4/5}.  \label{eq:Pi_BO}
\end{eqnarray}
The prefactors are derived using the scaling of large-scale quantities derived in the earlier sections.  The above model is called the BO phenomenology \cite{Bolgiano:JGR1959, Obukhov:DANS1959, Procaccia:PRL1989, Lohse:ARFM2010}. Note that $E_\theta(k) $ is called the entropy spectrum in literature.

If the effective buoyancy is active only at very low wavenumbers (as in Kolmogorov's phenomenology for fluid turbulence), then the convective turbulence can be approximated by Kolmogorov-like phenomenology, called the Kolmogorov-Obukhov (KO) model. This is akin to the turbulence phenomenology for passive scalar.  Here,  
\begin{eqnarray}
E_\theta(k) & \approx & \Pi_\theta [\Pi_u(k)]^{-1/3} k^{-5/3} \approx
\Delta^2 (\mathrm{Ra Pr})^{-0.13}   (kd)^{-5/3}. \\
E_u(k) & \approx & [\Pi_u(k)]^{2/3} k^{-5/3} \approx
d \left( \frac{\kappa}{d} \right)^2 (\mathrm{RaPr})^{0.88} (kd)^{-5/3}, \\
\Pi_u(k)  & \approx & \frac{1}{d} \left(\frac{\kappa}{d} \right)^3 (\mathrm{RaPr})^{3/2} C_{\epsilon_u} \approx \frac{1}{d} \left(\frac{\kappa}{d} \right)^3 (\mathrm{RaPr})^{1.3}, 
\label{eq:Pi_KO}
\end{eqnarray}

According to phenomenological model proposed by Procaccia and Zeitak~\cite{Procaccia:PRL1989} and L'vov~\cite{Lvov:PRL1991}, BO phenomenology should be active for small wavenumbers, while KO phenomenology for large wavenumbers. By matching the energy fluxes of the BO and KO phenomenologies [Eqs.~(\ref{eq:Pi_BO},\ref{eq:Pi_KO})], we can deduce the transition wavenumber called the Bolgiano wavenumber $k_{\mathrm{BO}}$, which is
\begin{equation}
k_{\mathrm{BO}} \approx \frac{1}{d} (\mathrm{RaPr})^{-0.05} \approx \frac{1}{d}.
\end{equation}
  Inverse of the Bolgiano wavenumber is called Bolgiano length.  Thus, according to Procaccia and Zeitak~\cite{Procaccia:PRL1989} and L'vov~\cite{Lvov:PRL1991}, BO scaling is expected for $k<k_{\mathrm{BO}}$, while KO scaling is expected for $k>k_{\mathrm{BO}}$.  From the above arguments, BO scaling may appear only for large aspect ratio boxes.

\subsection{Case 2: $\mathrm{Pe} \ll 1; \mathrm{Re} \gg 1$}

\begin{figure}
\includegraphics[scale=0.5]{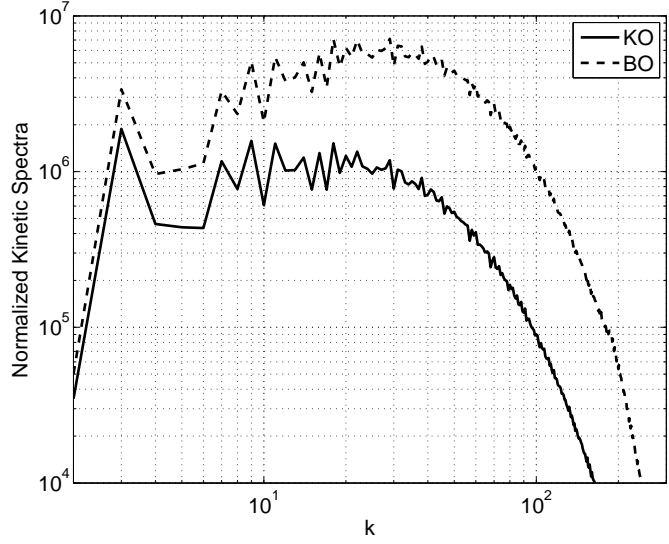}
\caption{Plot of normalized kinetic spectra $E^u(k)k^{5/3}$ (KO) and $E^u(k)k^{11/5}$ (BO) for Pr = 0 and Ra = $1.97 \times 10^4$. The spectrum is in good agreement with  the KO scaling. [Adopted from Mishra and Verma~\cite{Mishra:PRE2010}].}
\label{fig:pr0_spectra}
\end{figure}

Since $\Delta u_l/d \approx \kappa \theta_l/l^2$, the  entropy  spectrum is very steep.  Consequently, buoyancy is active only at very small wavenumbers.  Hence, Kolmogorov's theory of fluid turbulence is valid for this case, and we expect $k^{-5/3}$ spectrum for the velocity field.  Using the scaling described above
\begin{eqnarray}
E_u(k) & = & C_{\mathrm{Ko}} d \left(\frac{\nu}{d}\right)^2 \mathrm{Ra}^{2}(kd)^{-5/3}, \\
E_\theta(k) & = & d \Delta^2 (\mathrm{RaPr})^{2}(kd)^{-17/3}.
\end{eqnarray}

\begin{figure}
\includegraphics[scale=0.19]{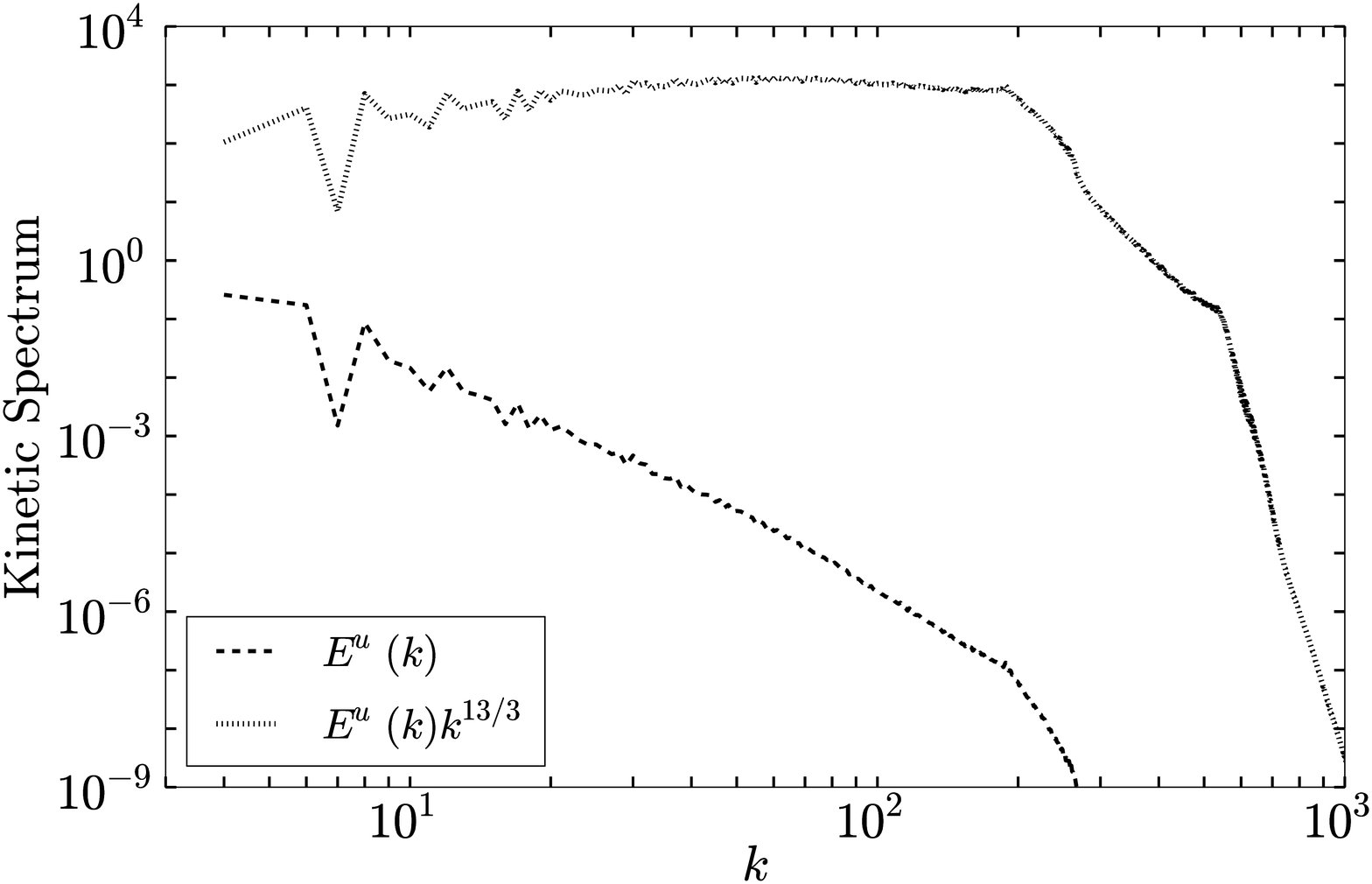}
\includegraphics[scale=0.19]{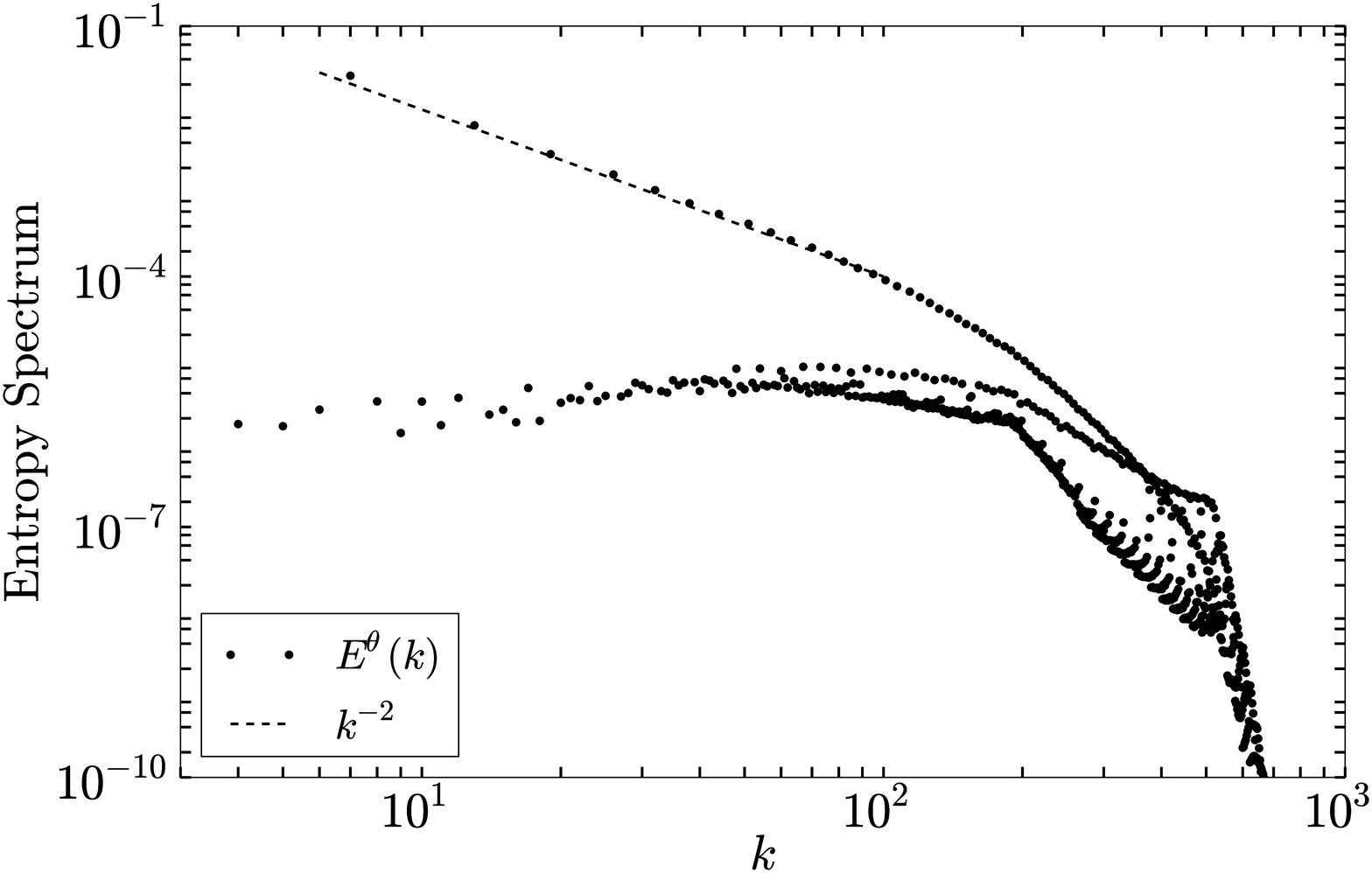}
\caption{Kinetic spectrum $E^u(k)$ (left) and entropy spectrum $E^{\theta}(k)$ (right) for Pr = $\infty$ and Ra = $10^8$. Normalized kinetic spectrum $E^u(k)k^{13/3}$ is flat, which is in good agreement with our predictions. Entropy spectrum exhibits dual branches. The upper branch corresponding to $\hat{\theta}(0,0,2n)$ modes fits well with $k^{-2}$ curve (dotted line)~\cite{Pandey:Preprint}.}
\label{fig:spectrum_infty}
\end{figure}

\subsection{Case 3: $\mathrm{Pe} \gg 1; \mathrm{Re} \ll 1$}
The momentum equation yields
\begin{equation}
\alpha g (\theta_\mathrm{res})_l \approx \frac{\nu u_l}{l^2}.
\end{equation}
We assume that  $(\theta_{\mathrm{res}})_l = \mathrm{Ra}^{-\delta} \theta_l$, and  a constant entropy flux $\epsilon_\theta = (\theta_L^2 u_L / L) C_{\epsilon_\theta} = (\theta_l^2 u_l / l) C_{\epsilon_\theta}$, which yields
\begin{eqnarray}
E_u(k) & = & d \left(\frac{\kappa}{d}\right)^2 \mathrm{Ra}^{2(1-\delta)}(kd)^{-13/3}, \\
E_\theta(k) & = & E_\theta(k)=d \Delta^2 (kd)^{-1/3}.
\end{eqnarray}

Measurement of energy and entropy spectra using laboratory experiment is quite difficult since the measurements are performed only at some of the points inside the container using thermal or velocity probes.  Strictly speaking, one cannot assume Taylor's hypothesis to relate frequency spectrum measured by the probes to the wavenumber spectrum.  In any case, the experimental results are inconclusive on the spectral exponents. 

On the numerical front, Borue and Orszag~\cite{Borue:JSC1997} reported the KO scaling for periodic box simulations.  We believe this is so because the box size of the simulation is of the order of Bolgiano length. Mishra and Verma~\cite{Mishra:PRE2010} performed numerical simulation for $\mathrm{Pr}=6.8$ and 1 for aspect ratio $2\sqrt{2}$, and observed inconsistency with both KO and BO scaling (see Fig.~\ref{fig:spectrum_pr6p8}),   Entropy spectrum, however, exhibits dual spectra because of the important role played by the $\hat{\theta}(0,0,2n)$ modes~\cite{Mishra:PRE2010}.  Thus the spectra for Case 1 remains inconclusive. We need higher resolution simulations for testing this regime.

Mishra and Verma~\cite{Mishra:PRE2010}  also performed simulations for  zero and small $\mathrm{Pr}$ and found good agreement with the scaling for Case 2, i.e., with the KO scaling~(see Fig.~\ref{fig:pr0_spectra}).  Regarding Case 3, the numerical results of Pandey \textit{et al.}~\cite{Pandey:Preprint} are in good agreement with our predictions (see Fig.~\ref{fig:spectrum_infty}).  

\section{7.~Large-scale modes and flow reversals}
Experiments on RBC reveal that in turbulent convection, the velocity near the lateral wall reverses randomly.   This phenomenon is called flow reversal.  Researchers have attempted to explain flow reversals using models involving stochastic resonance, plume dynamics, etc.~\cite{Ahlers:RMP2009}.  Recently Chandra and Verma~\cite{Chandra:PRE2011, Chandra:Arxiv2012} studied this phenomenon and showed that the flow reversals occur due to nonlinear interactions among the large-scale modes of the flow. In two-dimensional geometry, during the flow reversals, the corner rolls reconnect  and the vortices with same signs merge.  The newly formed vortex has flow direction opposite to the original vortex.   It will be interesting if the above reversal mechanism is at work in three-dimensional convection, as well as in dynamo.  The results of Chandra and Verma~\cite{Chandra:PRE2011, Chandra:Arxiv2012}  demonstrate that the large-scale modes play significant role in the flow reversal.


\section{8.~Conclusions}
In this paper we have presented scaling of large scale quantities, as well as the spectra and fluxes of velocity and temperature fields in convective turbulence.  The analyses have been performed using the scaling of bulk quantities.  The results presented in the paper are in good agreement with those obtained in earlier experiments and in numerical simulations, as well as with the predictions of GL theory.  These features demonstrate that bulk properties are very useful in determining characteristics of convection.   This is a significant step in modelling convective turbulence.  Boundary layer near the plates are important in convection since the energy to the bulk turbulence is supplied at the boundary layer. A detailed analysis combining the boundary layer and the bulk would be very useful is constructing a comprehensive understanding of convective turbulence.


\bibliographystyle{apsrev}
\bibliography{turbulence}

\end{document}